\documentclass[aps,prb,twocolumn,superscriptaddress,10pt]{revtex4-2}
\usepackage{graphicx}				
\usepackage{epsfig}
\usepackage{amsmath,amsthm} 
\usepackage{color}
\usepackage{verbatim}				
\usepackage{ulem}
\newcommand{\blue}[1]{{\color{black} #1}}

\begin{document}

\title{Orbital magnetization in the Nb-substituted Kagome metal CsV$_3$Sb$_5$}

\author{H.J. Elmers}\email{elmers@uni-mainz.de}
\affiliation{Institut f\"{u}r Physik, Johannes Gutenberg-Universit\"{a}t, Staudingerweg 7, D-55128 Mainz, Germany}

\author{O. Tkach}
\affiliation{Institut f\"{u}r Physik, Johannes Gutenberg-Universit\"{a}t, Staudingerweg 7, D-55128 Mainz, Germany}
\affiliation{Sumy State University, Kharkivska 116, 40007 Sumy, Ukraine}

\author{Y. Lytvynenko}
\affiliation{Institut f\"{u}r Physik, Johannes Gutenberg-Universit\"{a}t, Staudingerweg 7, D-55128 Mainz, Germany}
\affiliation{Institute of Magnetism of the NAS and MES of Ukraine, 03142 Kyiv, Ukraine}

\author{H. Agarwal}
\affiliation{Institut f\"{u}r Physik, Johannes Gutenberg-Universit\"{a}t, Staudingerweg 7, D-55128 Mainz, Germany}

\author{D. Biswas}
\affiliation{Diamond Light Source Ltd., Didcot OX11 0DE, United Kingdom}

\author{J. Liu}
\affiliation{Diamond Light Source Ltd., Didcot OX11 0DE, United Kingdom}

\author{A.-A. Haghighirad}
\affiliation{
Institute for Quantum Materials and Technologies, Karlsruhe Institute of Technology, Kaiserstr. 12, 76131 Karlsruhe, Germany}

\author{M. Merz}
\affiliation{
Institute for Quantum Materials and Technologies, Karlsruhe Institute of Technology, Kaiserstr. 12, 76131 Karlsruhe, Germany}
\affiliation{Karlsruhe Nano Micro Facility (KNMFi), Karlsruhe Institute of Technology, Kaiserstr. 12, 76131 Karlsruhe, Germany}

\author{S. Pakhira} 
\affiliation{
Institute for Quantum Materials and Technologies, Karlsruhe Institute of Technology, Kaiserstr. 12, 76131 Karlsruhe, Germany}

\author{G. Garbarino} 
\affiliation{ESRF, The European Synchrotron, 71 Avenue des Martyrs, CS40220, 38043 Grenoble Cedex 9, France}

\author{T.-L.\,\,Lee}
\affiliation{Diamond Light Source Ltd., Didcot OX11 0DE, United Kingdom}

\author{J. Demsar}
\affiliation{Institut f\"{u}r Physik, Johannes Gutenberg-Universit\"{a}t, Staudingerweg 7, D-55128 Mainz, Germany}

\author{G. Sch{\"o}nhense}
\affiliation{Institut f\"{u}r Physik, Johannes Gutenberg-Universit\"{a}t, Staudingerweg 7, D-55128 Mainz, Germany}

\author{M. Le Tacon}
\affiliation{
Institute for Quantum Materials and Technologies, Karlsruhe Institute of Technology, Kaiserstr. 12, 76131 Karlsruhe, Germany}

\author{O. Fedchenko}
\affiliation{Physikalisches Institut, Goethe Universit{\"a}t Frankfurt,
Max-von-Laue-Str. 1, D-60438 Frankfurt am Main
}

\date{October 2025}

\begin{abstract}

This study uses angle-resolved photoemission spectroscopy to examine the low-temperature electronic structure of Cs(V$_{0.95}$Nb$_{0.05}$)$_3$Sb$_5$, demonstrating that partially substituting V atoms with isoelectronic Nb atoms results in \blue{an increase of the band width} and enhanced gap opening at the Dirac-like crossings due to the resulting chemical pressure. This increases the magnetic circular dichroism  signal in the angular distribution (MCDAD) compared to CsV$_3$Sb$_5$, enabling detailed analysis of magnetic circular dichroism in several bands near the Fermi level. These results \blue{substantiate} the predicted coupling of orbital magnetic moments to three van Hove singularities near the Fermi level at M points. Previous studies have observed that Nb doping \blue{lowers the charge density transition temperature} and increases the critical temperature for superconductivity. This article demonstrates that Nb doping concomitantly increases the magnetic circular dichroism signal attributed to orbital moments.

\end{abstract}

\maketitle

\section{Introduction}

The correlation between spontaneous electronic ordering
and spatial electronic modulations, including magnetism, chirality, charge density waves (CDWs), and unconventional superconductivity, is one of the key challenges to be solved in contemporary condensed matter research~\cite{Neupert2021,Agterberg2020,Fernandes2019}.
In this field of research, the layered canonical kagome compounds AV$_3$Sb$_5$ (A = K, Cs, Rb)~\cite{Ortiz2020} have recently emerged as an interesting class of materials.
Their electronic structure is characterized by the presence of flat
bands, multiple van Hove singularities (vHSs), Dirac cones, and
non-trivial band topology.
These materials therefore provide a unique platform for exploring novel electronic states of matter with intertwined orders~\cite{Neupert2021}. 
Of particular interest is the electronic superlattice observed at low temperatures~\cite{Ortiz2020},
which is commonly associated with a CDW instability.

One of the most fascinating theory predictions is a spontaneous breaking of the time-reversal symmetry in the low-temperature CDW state of CsV$_3$Sb$_5$~\cite{Feng2021,Chen2023}.
Indeed, this prediction has been confirmed by a number of experiments~\cite{Jiang2021,Wang2021,Wei2024,Zhou2022,Yu2021,Yang2020,Mielke2022,Khasanov2022,Kautzsch2023,Elmers2025,gui2025}, though other studies have produced conflicting results~\cite{Farhang2023,Wang2024,Saykin2023,Li2022}, such as negligible spontaneous magneto-optical Kerr effect~\cite{Saykin2023}.

\begin{figure}
\includegraphics[width=\columnwidth]{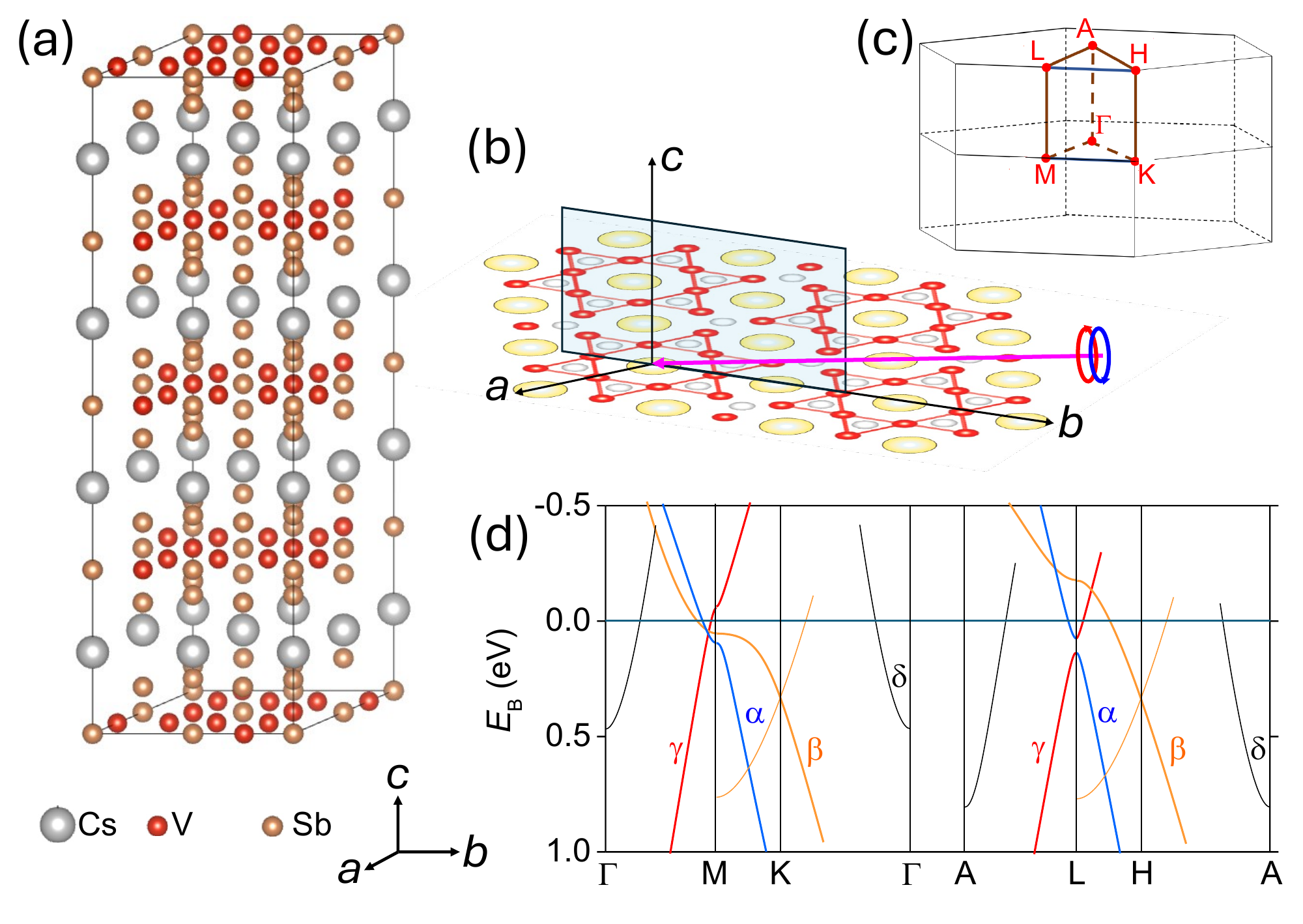}
\caption{\label{Fig1} 
(a) Crystal structure of the canonical kagome metal CsV$_3$Sb$_5$ at low temperatures, indicating the orthorhombic 2$\times$2$\times$4 superstructure~\cite{Kautzsch2023}. (b) Experimental ARPES geometry. The angle of incidence of the circularly polarized X-rays is 22.5$^\circ$ with respect to the sample a-b plane. (c) Brillouin zone of the pristine crystal structure in the high temperature phase (we use the high-temperature phase notation throughout the paper).
(d) Schematic band structure of pristine CsV$_3$Sb$_5$ highlighting the electronic bands derived from the V 3$d$-orbitals forming the three van Hove singularities near the Fermi level at the M-points. 
}
\end{figure}

Understanding the origin of the CDW and the nature of
the interplay between electronic and lattice degrees of
freedom in CsV$_3$Sb$_5$ therefore require further
studies~\cite{Li2021,Subires2023,Zhong2023,Zhong2023a,Zhong2024}.
Here, pressure tuning provides a reversible method of adjusting the balance
between competing energy scales allowing the study of complex phase diagrams in these materials without introducing chemical disorder~\cite{Stier2024,Wenzel2023}. Indeed, a recent diffraction study of CsV$_3$Sb$_5$  shows that the CDW order undergoes a transformation from a 2$\times$2 superstructure to a fractional $3/8$ CDW wave vector at higher pressures, challenging the Peierls-like nesting scenarios~\cite{Stier2024}.

Instead of tuning the electronic properties with hydrostatic pressure, chemical pressure via the partial substitution of atoms with isoelectronic atoms of different sizes can also be used to tune the electronic structure, 
and is compatible with band structure measurements using angle-resolved photoemission spectroscopy (ARPES).
Indeed, a recent doping-dependent study~\cite{Oey2022} demonstrated 
that the substitution of Sb by Sn destabilizes the $2\times 2$ superlattice, leading to the formation of a superlattice similar to that observed in the pressure-dependent study~\cite{Stier2024} at low temperatures.
However, the doping-induced phase can exhibit shorter and highly anisotropic correlation lengths within the $a-b$ layers. 
On the other hand, Nb substitution has been shown to lower the CDW transition temperature and render the CDW superstructure two-dimensional~\cite{Xiao2023a}, while increasing the critical temperature for superconductivity~\cite{Zhou2023}.

Qualitatively, both doping and chemical or mechanical pressure affect the ratio of the out-of-plane
and in-plane lattice parameters $c/a$, which has been shown
to play a critical role in controlling the electronic properties~\cite{Frachet2024,Ritz2023}. 

In this article, we present results of a low-temperature ARPES study 
of the compound Cs(V$_{1-x}$Nb$_{x})_3$Sb$_5$ ($x=0.05$). In this kagome metal, the V atoms - which are assumed to carry the loop current order and the orbital magnetic moments
via their $3d$ orbitals - are partially substituted by the larger, isoelectronic, Nb atoms. We demonstrate that chemical pressure increases the average orbital overlap of the V/Nb $3d/4d$ orbitals. This leads to \blue{an increase of the band width}, enhanced gap opening at the Dirac-like crossings, and increased magnetic circular dichroism. The latter allows a detailed view of the circular dichroism of several bands near the Fermi level. Our analysis \blue{substantiates} the predicted coupling of the orbital magnetic moments with the three van Hove singularities near the Fermi level at the M points.

\section{Experimental}

The experiments were performed on single crystals of Cs(V$_{1-x}$Nb$_{x})_3$Sb$_5$ ($x=0.05$) grown by the flux method. The crystals were characterized using X-ray diffraction and energy dispersive X-ray spectroscopy~\cite{Frachet2024,Stier2024}.
Above the charge-density-wave transition temperature of $60$~K, Nb-substituted CsV$_3$Sb$_5$ has the same structure as the pristine compound. 
High-resolution X-ray diffraction measurements were performed at 22~K on the same sample used for the photoemission experiments at the ID15B beamline of the ESRF (France). A monochromatic X-ray beam with an energy of 30.17~keV was used, focused down to 2$\times$4~$\mu$m$^2$ at the sample position. The structural properties are consistent with previous results~\cite{Stier2024}. The Nb concentration  $x$ of niobium was determined by energy dispersive X-ray spectroscopy to be $x=0.052(1)$.



Circular dichroism experiments in the soft x-ray range were performed 
at the soft x-ray ARPES endstation of Beamline I09 at the Diamond Light Source, 
UK~\cite{Schmitt2024}. All photoemission experiments were conducted at 30~K.
\blue{The single crystals were freshly cleaved in ultrahigh vacuum. No magnetic field was applied before or during the experiment.
The plane of incidence of the circularly polarized X-rays coincides with the in-plane $b$-axis and with the surface normal along the $c$-axis [see Fig.~\ref{Fig1}(b)]. 
The degree of circular dichroism of the incident X-rays was greater than 95\%. The angle of incidence for the circularly polarized X-rays was $22.5^0$ with respect to the sample surface, which was oriented to align the $\Gamma$-M-L plane with the incident beam [see Fig.~\ref{Fig1}(b,c)]. The total energy resolution was set to 50~meV. The spot size at the sample was limited to $50\times50$~$\mu$m$^2$ by the exit slits. We optimized the probed area to maximize the contrast of the valence band structure.}


\section{Results}

We begin by discussing the parent crystal structure of pristine CsV$_3$Sb$_5$ as 
shown in Fig.~\ref{Fig1}a. CsV$_3$Sb$_5$ consists of a planar arrangement of V atoms forming a kagome lattice consisting of three sets of parallel V atom lines (red), with Sb atoms (orange) as nearest neighbors. The V-Sb planes are separated by a layer of Cs atoms (gray). This structure gives rise to a two-dimensional electronic structure because the valence bands are formed by V and Sb orbitals.

Figure~\ref{Fig1}c shows the Brillouin zone and high symmetry points for the high temperature hexagonal phase (we use the high-temperature Brillouin zone also when discussing the electronic structure of the low-temperature phase), while
Figure~\ref{Fig1}d presents the schematic band structure as described by Kang et al.~\cite{Kang2022}. 
At the M-point, three van Hove singularities near the Fermi level are derived from V $3d$ orbitals and are labeled as $\alpha$, $\beta$, and $\gamma$ band. Here, only the van Hove singularity of the $\beta$ band exhibits dispersion along $k_z$. The electron-like $\delta$ band, which has a minimum at the $\Gamma$-point, is derived from the Sb $s/p/d$ orbitals and shows pronounced dispersion from the $\Gamma$-point to the A-point.
 As Nb is isoelectric to V, we expect that 5\% of V atoms are substituted by Nb atoms.

\begin{figure}
\includegraphics[width=\columnwidth]{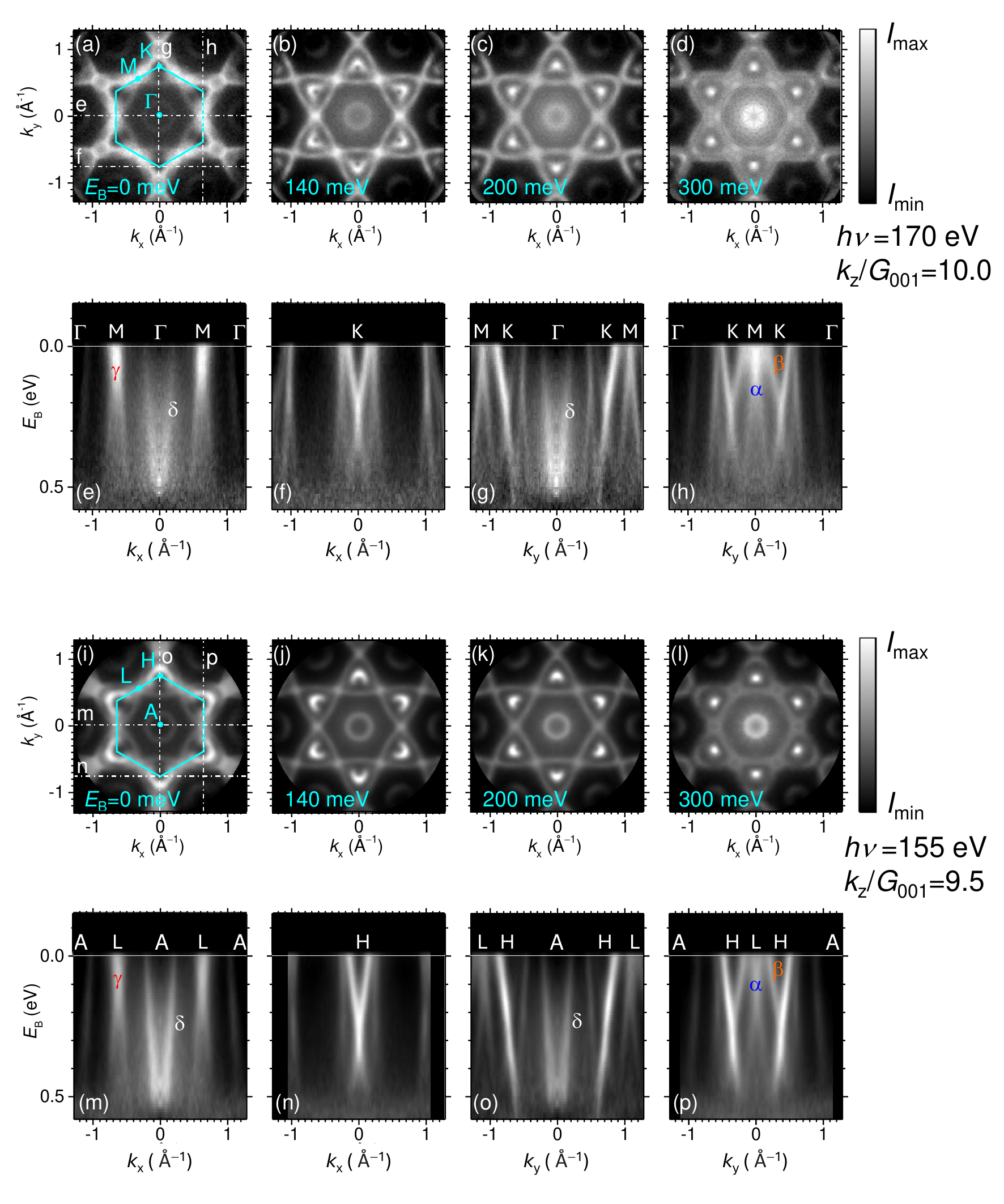}
\caption{\label{FigInt} 
(a-d) Constant energy cuts of the photoemission intensities in the $k_x - k_y$ plane at the indicated binding energies, measured at a photon energy of 170~eV for Nb-substituted CsV$_3$Sb$_5$ (probing the $\Gamma$-K-M plane - see Fig.~\ref{Fig1}). 
(e-h) Band dispersions along the indicated high symmetry directions in reciprocal space.
(i-p) Similar data measured at a photon energy of 155~eV (probing the A-L-H plane - see Fig.~\ref{Fig1}).
}
\end{figure}

\begin{figure}
\includegraphics[width=\columnwidth]{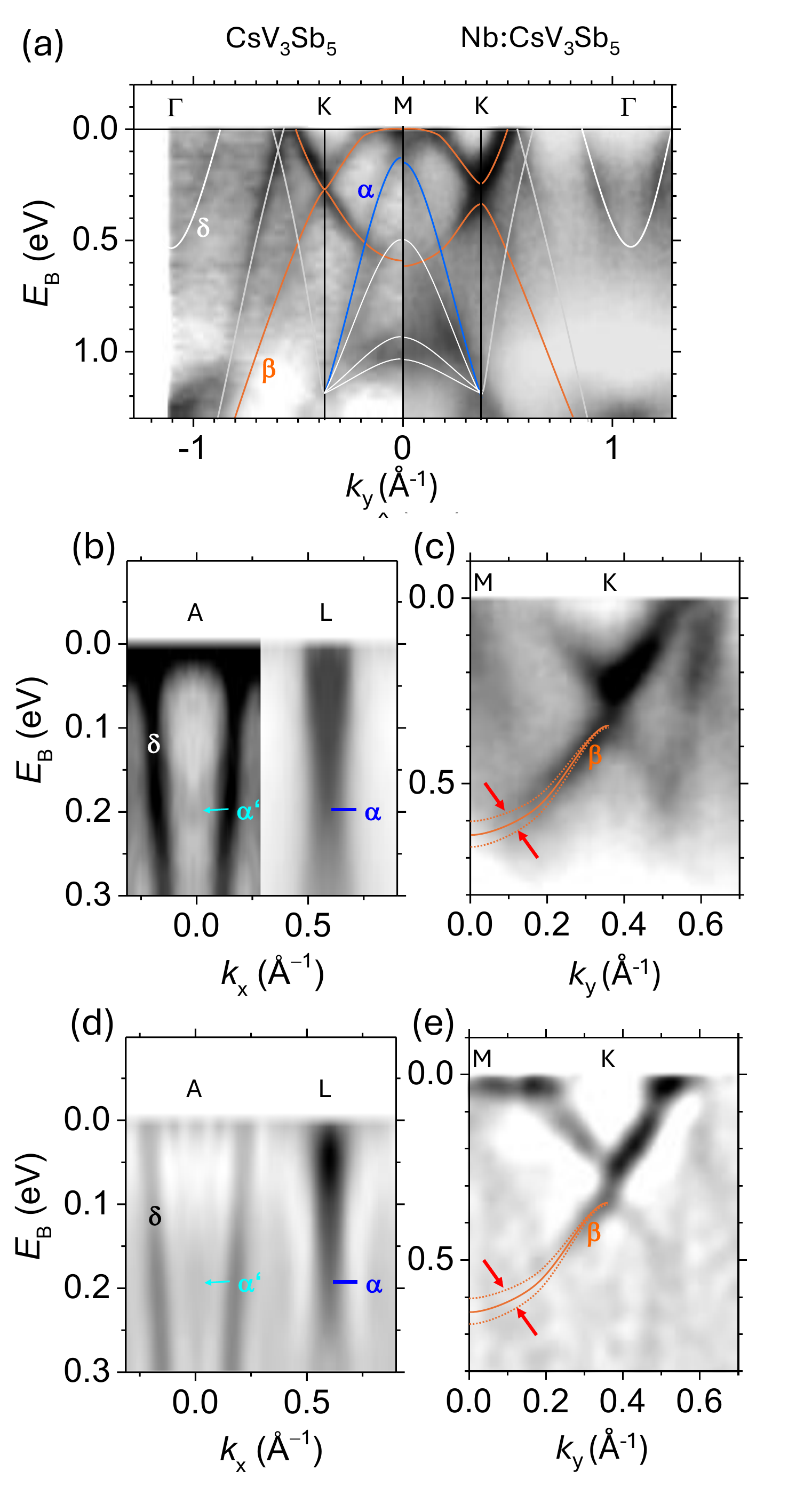}
\caption{\label{FigComp} 
(a) Comparison of the band dispersions of pristine CsV$_3$Sb$_5$ and 
Nb-substituted CsV$_3$Sb$_5$ at $T=30$~K. Data for pristine CsV$_3$Sb$_5$ have been measured at a photon energy of 250~eV and the data for the Nb-substituted compound at 170~eV, both corresponding to a cut through the $\Gamma$ point along $k_z$.
(b) Band dispersion along the A - L direction for Nb-substituted CsV$_3$Sb$_5$ at 30~K. The adapted intensity scale 
was changed at $k_x=0.3$~\AA$^{-1}$ to show the backfolded band $\alpha'$. 
(c) Band dispersion along the M - K direction for Nb-substituted CsV$_3$Sb$_5$ at 30~K.
\blue{
(d) Laplacian derivative of the band dispersion along the A-L direction using the data shown in (b). 
(e) Laplacian derivative of the data shown in (c).
}
}
\end{figure}

\subsection{Electronic structure of Cs(V$_{0.95}$Nb$_{0.05}$)$_3$Sb$_5$}

Below 60~K, both the pristine sample and the Nb-substituted CsV$_3$Sb$_5$ sample exhibit a CDW reconstruction. As mentioned, substituting V atoms with larger Nb atoms is expected to strengthen the orbital interaction. 
In order to investigate how the Nb substitution influences the electronic states in three-dimensional momentum space, we performed photoemission experiments in the soft X-ray regime, which enabled us to vary the probed perpendicular (k$_z$) momentum.

Soft X-ray photoemission experiments result in photoemission intensity distributions $I(E_B,k_x,k_y)$, which are binned in a three-dimensional array. The data shown in Fig.~\ref{FigInt} were measured simultaneously in the hybrid time-of-flight and dispersive analyzer mode~\cite{Schmitt2024,Schoenhense2020}. 
This mode allows the simultaneous recording of an energy interval of 0.6~eV with an energy resolution of 50~meV. 
To address the open question of spontaneous magnetic order in the CDW phase of Nb-substituted CsV$_3$Sb$_5$, we used left and right circularly polarized soft X-rays at photon energies of 170~eV and 155~eV, corresponding to $k_z=10.0G_{001}$ (probing the $\Gamma$-K-M plane) and $k_z=9.5G_{001}$ (probing the A-L-H plane), respectively. When discussing the electronic structure, the data recorded for the two circular polarizations were averaged and symmetrized (see Ref.~\onlinecite{Elmers2025}).

Figures~\ref{FigInt}a-d present the measured Fermi surface and constant energy sections at higher binding energy $E_B$ in the $\Gamma$-K-M plane of the Brillouin zone recorded at 30~K, \textit{i.e.}, at $T<T_{\rm CDW} $. 
The central ring corresponds to the Sb $s/p/d$ orbital-derived $\delta$ band. 
The $\beta$ and $\gamma$ bands merge at the M point, resulting in a high photoemission intensity at the Fermi energy, $E_F$.
The $\gamma$ band surrounds an area of low photoemission intensity around the K point.
At higher binding energies (Fig.~\ref{FigInt}b-d) the $\gamma$ band forms a triangularly distorted feature centered at the K-point that shrinks to a single point of high intensity at about $E_B=300$~meV.
At binding energies beyond 300~meV the band opens again, indicating a behavior similar to that of a Dirac state.

The sections along the $\Gamma$-K-M directions (Fig.~\ref{FigInt}g,h) show the hole-like dispersion of the $\alpha$ and $\beta$ bands. These bands form the van Hove singularities (vHS1 and vHS2)  near the Fermi energy at the M-point.
The section along the $\Gamma$-M direction (Fig.~\ref{FigInt}e)
reveals the electron-like dispersion of the $\alpha$ band, while the top of the $\beta$ band is shifted to above $E_F$. In this section, the hole-like dispersion of the $\gamma$ band is also observed, which is part of vHS3 below $E_F$.
Here, the opposite curvatures form the saddle point of the van Hove singularity.
Figure~\ref{FigInt}f shows the cross-section through the Dirac-like state at the K-point, revealing the nearly linear dispersion near $E_B=0.3$~eV. 

Similar band dispersions are observed for 155~eV photon excitation (Fig.~\ref{FigInt}i-p), with the perpendicular momentum $k_z$ corresponding to $9.5 G_{001}$. Consequently, the Brillouin zone is probed in the A-H-L plane. The first notable difference between the two $k_z$ values 
is that the $\beta$ band near $E_F$ clearly crosses the Fermi level near the L point with a linear dispersion, whereas the band shows a small negative curvature near the M-point  (Fig.~\ref{FigInt}h,p).
This observation is in agreement with the calculated results for the pristine compound. 
The second difference is the higher binding energy of the $\alpha$ band minimum, which appears at $E_B=0.55$~eV for 170~eV and below 0.6~eV for 155~eV. This is also in agreement with theoretical results~\cite{Wu2024,Hu2022b,Kang2022}
and previous photoemission results obtained for the pristine
CsV$_3$Sb$_5$ kagome metal~\cite{Hu2022c,Kang2022,Luo2022,Lou2022,Elmers2025}. 

\subsection{Comparison of electronic structures of Cs(V$_{0.95}$Nb$_{0.05}$)$_3$Sb$_5$ and CsV$_3$Sb$_5$}

Figure~\ref{FigComp}a shows a direct comparison of the photoemission intensity for pristine CsV$_3$Sb$_5$ and Nb-substituted CsV$_3$Sb$_5$ at $T=30$~K.
These measurements were performed in pure hemispherical analyzer mode at a lower pass energy, where constant energy sections are recorded sequentially. The total energy resolution was set to 30~meV.

The first notable result is the observation of sharp band features for bands derived from V orbitals, despite the partial substitution of V atoms by Nb atoms. Although broadening of the band features was expected as a result of disorder~\cite{Zhou2023}, the coherent Bloch waves appear to be unaffected by scattering at the larger Nb atoms. Furthermore, only minor variations in the band structure are evident when comparing the pristine compound with the Nb-substituted one. 

The most obvious difference is the increased band gap at the K point for the Nb-substituted case, where the Dirac-like crossing of the $\beta$ occurs.
The band gap is approximately 100~meV. For the pristine compound, we found a nearly vanishing band gap, resulting in a Dirac-like dispersion, which is in agreement with previous findings~\cite{Kang2022}.
The photoemission intensity does not vanish within the band gap because the gap is caused by spin-orbit coupling, which avoids band crossing and results in typical band broadening with finite photoemission intensity within the small energy gap.

\begin{figure}
\includegraphics[width=\columnwidth]{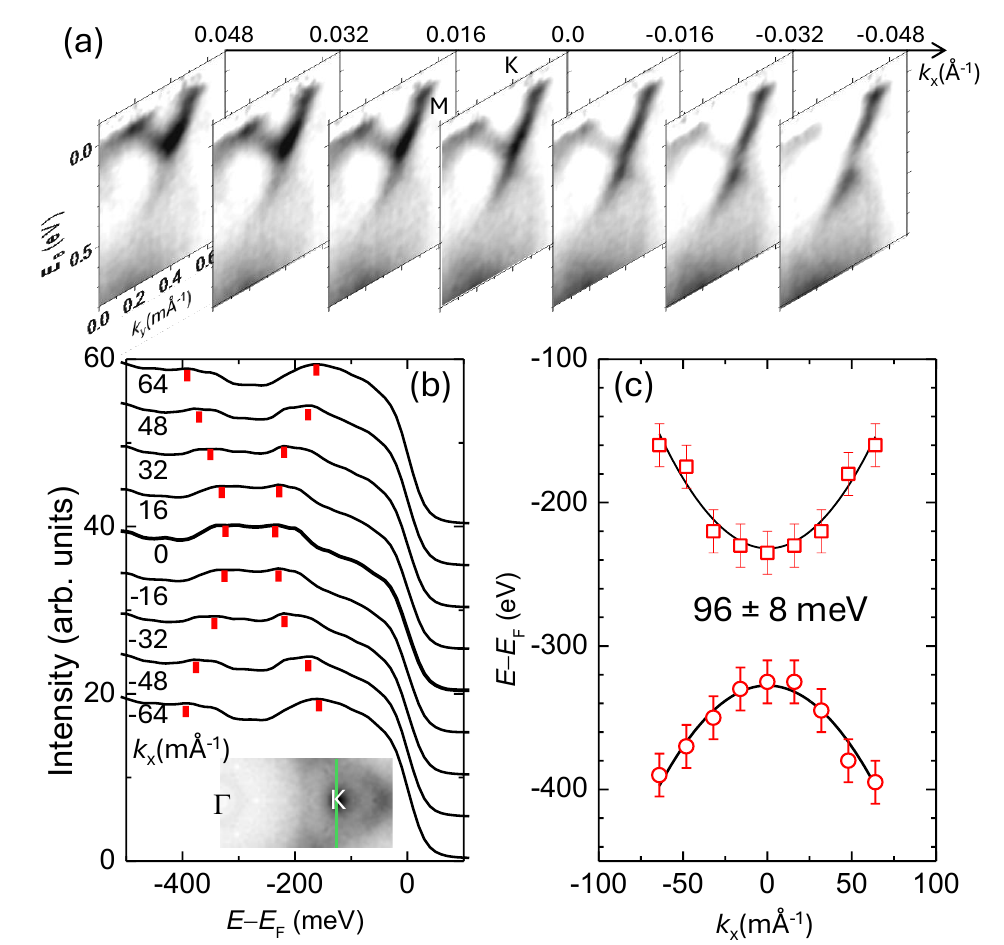}
\caption{\label{FigDiracGap} 
\blue{(a) Sections of the photoemission intensity, $I(E_B,k_y)$, at equidistant $k_x$ values parallel to the K - M direction in the vicinity of the gap at the Dirac point from the same data set shown in Fig.~\ref{FigComp}(c). 
(b) Set of energy distribution curves 
for equidistant momentum values along the profile indicated in the inset. The red marks indicate the maximum intensities near the Dirac point at $E_B=0.3$~eV. 
(c) Binding energies of the maximum intensities at lower (squares) and higher (circles) binding energy as a function of $k_y$. Parabolic fits (full lines) to the binding energies result in an energy gap of $(96\pm 8)$~meV.}}
\end{figure}

\blue{To study the gap opening at the Dirac point in more detail, Fig.~\ref{FigDiracGap}(a) shows a series of $I(E_B,k_y)$ sections that visualize the $I(E_B,k_x,k_y)$ data and define the position of the K point along the M - K direction. Energy distribution curves [Fig.~\ref{FigDiracGap}(b)] along the direction perpendicular to the $\Gamma$ - K direction reveal the dispersion of the bands near the Dirac point. The gap opening results from parabolic fits to the dispersion in the vicinity of the Dirac point [see Fig.~\ref{FigDiracGap}(c)].}

An additional visible difference is the increased bandwidth of the $\beta$ band. The maximum energy appears to be fixed at the Fermi level, likely due to the high density of states associated with the van Hove singularity. Meanwhile, the minimum energy at the M point has shifted to higher binding energies by 50~meV in the Nb-substituted compound. Additionally, the second Dirac point between the $\Gamma$ and K points shifts to higher binding energies with Nb substitution. As these bands are related to V $3d$ states with $d_{xz}/d_{yz}$ orbital character~\cite{Kang2022} we attribute the \blue{increase of the band width} to the increased orbital overlap caused by partial substitution of V atoms with the larger Nb atoms.

\blue{For pristine CsV$_3$Sb$_5$, a doubling of the $\beta$ band at $E_F$ near the M point has been reported~\cite{Kang2022a}. 
This doubling has been observed across the entire AV$_3$Sb$_5$ (A = K, Rb, Cs) family in the low temperature CDW state~\cite{Kang2022a}. However, it is difficult to observe this effect in our experimental data on the pristine compound.
In contrast, a close inspection of the data for the Nb-doped variant reveals the doubling, as indicated by the red arrows in
Fig.~\ref{FigComp}(c)}.
This band doubling can be understood as a consequence of three-dimensional charge ordering, where the modulation along the out-of-plane direction results in folding the Brillouin zone along the $k_z$ direction. This superimposes the $k_z = \pi/2$ bands on the $k_z = 0$ bands~\cite{Kang2022a}.

It has been suggested that this splitting is due to a particular three-dimensional reconstruction of the kagome lattice of V ions, where alternating Star-of-David and tri-hexagonal distortions occur along the $c$-axis. In addition, trilinear coupling of M and L phonon modes lead to instabilities, resulting in further distortions along the $c$-axis~\cite{Kang2022a}. Therefore, it is likely that the C$_6$ rotational symmetry is broken, similar to the case of pristine CsV$_3$Sb$_5$~\cite{Elmers2025}.

Additionally, in the case of Nb-substituted CsV$_3$Sb$_5$, we observe the back-folding of the $\alpha$ band from the L point to the A point, indicated by $\alpha'$ (Fig.~\ref{FigComp}b). This back-folding is indicative of the in-plane 2$\times$2 reconstruction due to charge ordering, which has also been observed in the pristine CsV$_3$Sb$_5$ compound~\cite{Jiang2023,Luo2022a}.


We conclude that in Nb-substituted CsV$_3$Sb$_5$ the structural distortions due to charge ordering are similar to those observed in pristine CsV$_3$Sb$_5$, but different to those observed in KV$_3$Sb$_5$ and RbV$_3$Sb$_5$.
The changes in the band structure caused by the chemical pressure resulting from Nb substitution at V sites can be qualitatively explained by an increased band overlap and by an increased spin-orbit coupling. These changes occur in the bands derived from V/Nb $3d/4d$ orbitals. In contrast to a recent theoretical study of pressure-dependent band structure effects~\cite{Wenzel2023}, the band states localized on Sb atoms appear to be less affected.

\subsection{Magnetic circular dichroism in Nb-substituted CsV$_3$Sb$_5$}

\begin{figure*}
\includegraphics[width=0.7\textwidth]{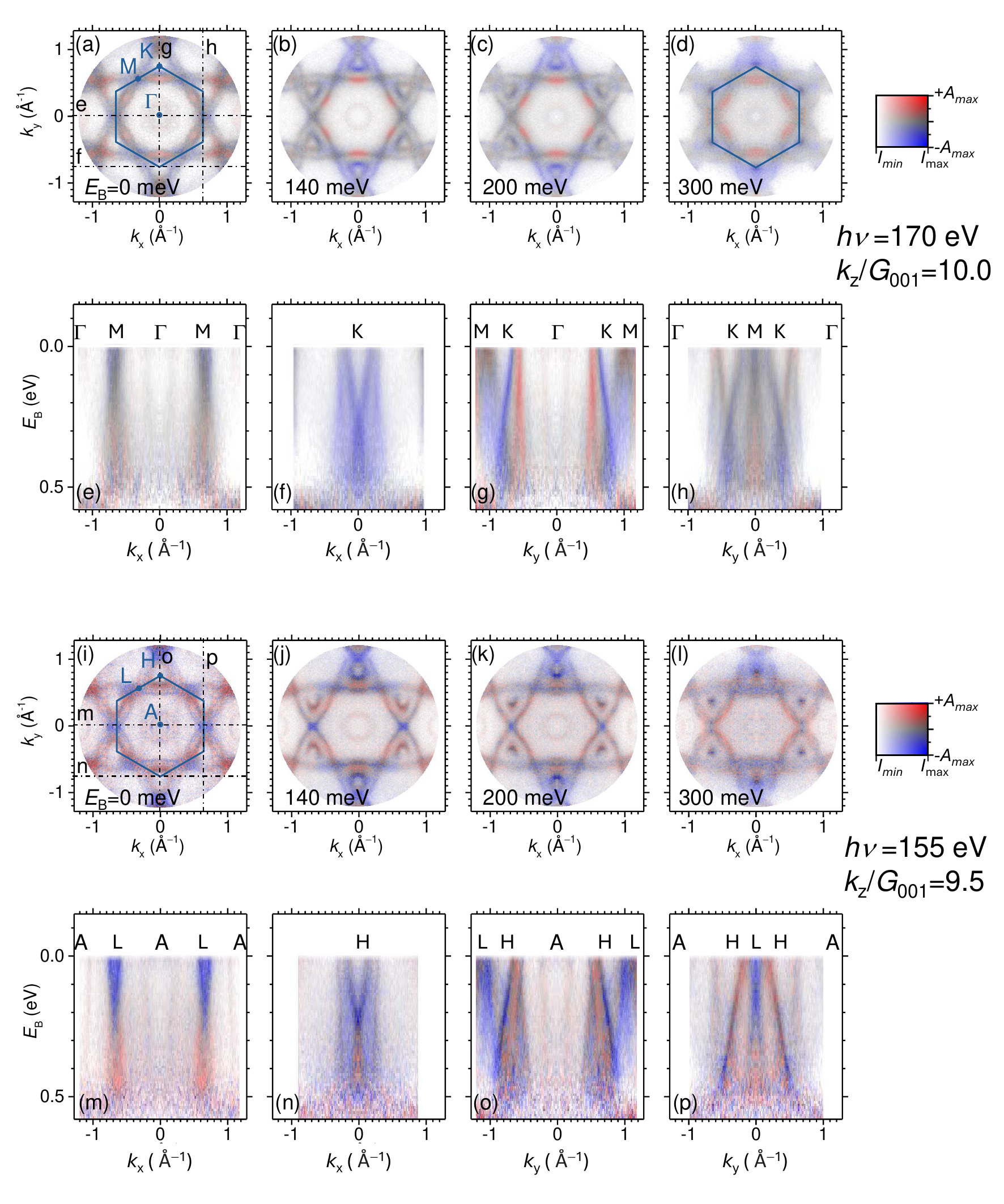}
\caption{\label{FigMCD} 
(a-d) Constant energy sections of the magnetic circular dichroism (MCDAD) and photoemission intensities in the $k_x - k_y$ plane at the indicated binding energies, measured at a photon energy of 170~eV for Nb-substituted CsV$_3$Sb$_5$. 
The photoemission intensity is overlaid with the MCDAD asymmetry in a
two-dimensional color scale as indicated on the right.
The maximum asymmetry of the color scale (right) is $A_{\rm max}=0.1$.
(e-h) Band dispersion of the MCDAD asymmetry along the indicated lines in panel (a).
(i-p) Similar data measured at a photon energy of 155~eV.
}
\end{figure*}

To investigate the potential for spontaneous time-reversal symmetry breaking in Nb-substituted CsV$_3$Sb$_5$, we conducted magnetic circular dichroism studies, adopting the methodology outlined in Ref.~\onlinecite{Elmers2025}. First, we determined the asymmetry in the photoemission spectra measured with X-rays of left circular polarization, $I^{-}$, and right circular polarization, $I^{+}$ , giving the asymmetry signal $A^*=(I^+-I^-)/(I^++I^-)$.

In the present measurement geometry, in which the scattering plane of the X-rays coincides with the mirror plane of the Brillouin zone (see Fig.~\ref{Fig1}b), this asymmetry comprises two contributions. 
The first is the non-relativistic circular dichroism in the angular distribution (CDAD), a geometric effect that is strictly antisymmetric with respect to the photon plane of incidence~\cite{Westphal1989}. The second is the magnetic circular dichroism in the angular distribution (MCDAD), which is strictly symmetric with respect to the scattering plane of the X-rays.
To separate the two contributions, we calculate the CDAD using $A_{\rm CDAD}(k_x,k_y)=(1/2)[A^*(k_x,k_y)-A^*(k_x,-k_y)]$ and the magnetic circular dichroism via  
$A_{\rm MCDAD}(k_x,k_y)=(1/2)[A^*(k_x,k_y)+A^*(k_x,-k_y)]$. A detailed discussion of the separation procedure can be found in Ref.~\onlinecite{Fedchenko2024}.

Note that the time-reversal symmetry prohibits the occurrence of $A_{\rm MCDAD}$.
Therefore, significant finite values of the MCDAD are only expected to be observed in systems with spontaneous magnetic order. Reports of a natural magnetic circular dichroism occurring in photoemission in chiral systems are analogous to absorption effects known in biochemistry~\cite{Kelly2005}.
However, these effects are thought to be very small.

By contrast, our experimental data reveal $A_{\rm MCDAD}$ values of around 0.1 (see Fig.~\ref{FigMCD}), which are symmetric with respect to $k_y=0$.
Here, the MCDAD asymmetry is plotted on a two-dimensional color scale to also highlight the band dispersions.
Prominent negative (blue) values of $A_{\rm MCDAD}$ appear near the M- and L-points in Fig.~\ref{FigMCD}a-d,i-l for both $k_z=0$ ($\Gamma$-K-M plane) and $k_z=0.5G_{001}$ (A-L-H plane), respectively. Positive (red) $A_{\rm MCDAD}$ values are observed along the directions from the $\Gamma$ point to the K point and from the A point to the H point, respectively.
The MCDAD asymmetry is restricted to the bands associated with the V 3$d$ orbitals.
The electron-like $\alpha$ band, dispersing along the M-$\Gamma$ direction (Fig.~\ref{FigMCD}e) shows a negative MCDAD asymmetry, similar to that of the corresponding band along the A-L direction (Fig.~\ref{FigMCD}m).
This negative MCDAD asymmetry also dominates the hole-like $\alpha$ band dispersing along  
the M-K and L-H directions, as shown in Figs.~\ref{FigMCD}g,h,o,p.
In contrast, the $\beta$ band dispersion along the M-K and L-H directions exhibits a positive MCDAD asymmetry.

The most significant MCDAD asymmetry is observed near the two M and L points at $k_y=0$ (see Fig.~\ref{FigMCD}a,i). 
At these points, the MCDAD symmetry axis coincides with the CDAD antisymmetry axis, where the CDAD vanishes. 

\begin{figure}
\includegraphics[width=\columnwidth]{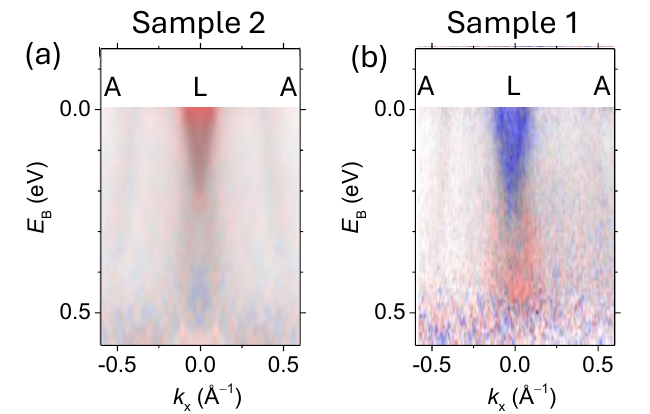}
\caption{\label{FigMreversal} 
(a) Band dispersion of the MCDAD asymmetry along the A - L line (similar as in Fig.~\ref{FigMCD}m) for a second sample on the same sample holder measured at a photon energy of 155~eV at 30~K.
(b) For comparison data from Fig.~\ref{FigMCD}m are shown on the same scale. 
}
\end{figure}

If the MCDAD asymmetry is indeed the result of spontaneous time-reversal symmetry breaking, the sign of the momentum- and energy-dependent MCDAD signal is expected to flip between different domains or different crystals. Indeed, the data recorded on two small Nb-substituted CsV$_3$Sb$_5$ single crystals, mounted on the same sample holder, 
display a change  in the sign of $A_{\rm MCDAD}$. Fig.~\ref{FigMreversal} presents the MCDAD signals near the M-points at $k_y=0$ of the two samples. Here, the $\alpha$ band exhibits a positive $A_{\rm MCDAD}$ value in the first crystal and a negative value in the second crystal. The hole-like dispersing $\gamma$ band shows the opposite behavior. This observation \blue{strongly suggests} that the observed MCDAD reflects the orbital magnetic order in CsV$_3$Sb$_5$.

\subsection{Detailed mapping of the magnetic circular dichroism in Nb-substituted CsV$_3$Sb$_5$}

\begin{figure*}
\includegraphics[width=0.7\textwidth]{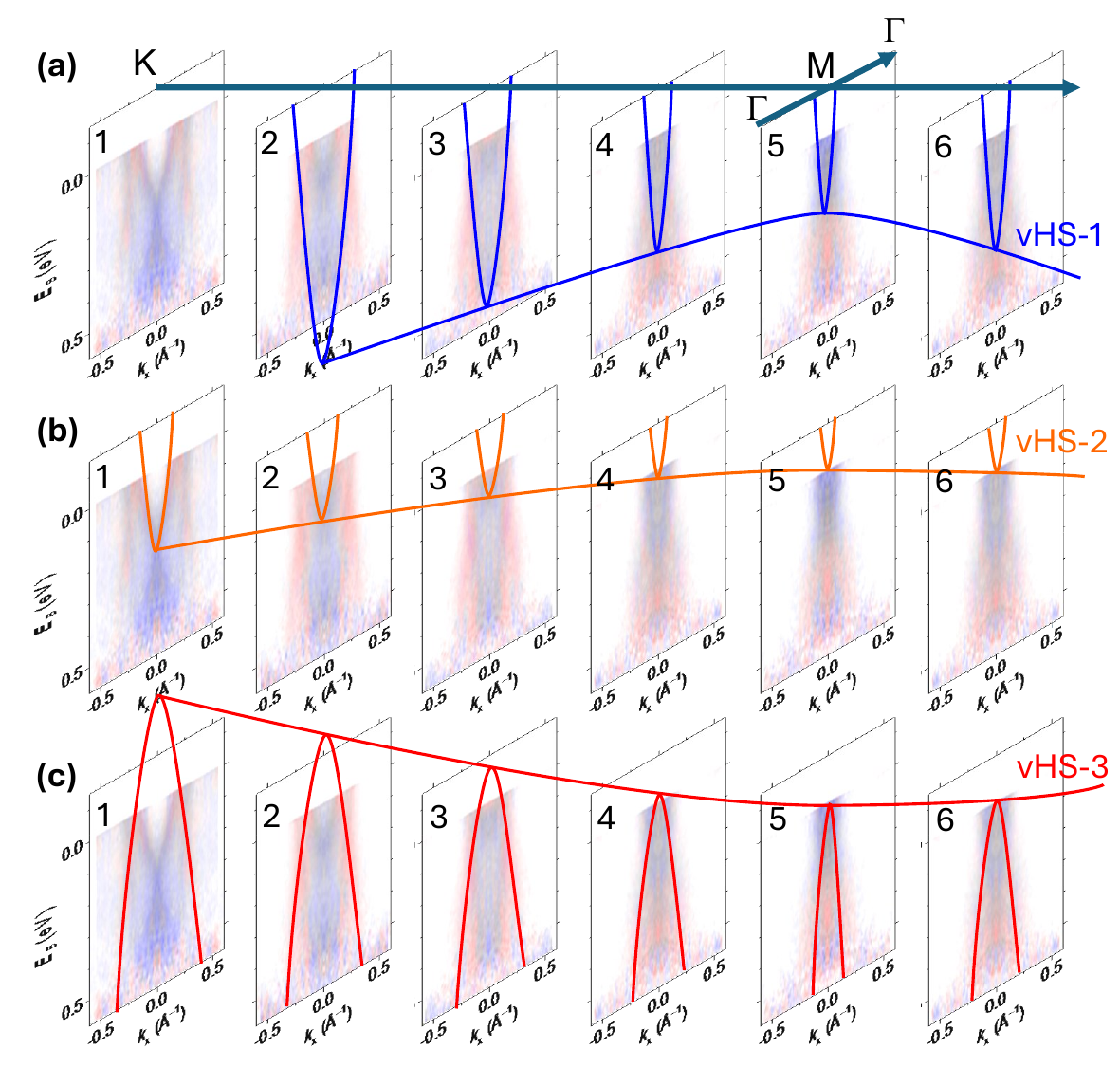}
\caption{\label{FigvHs1} 
Mapping of multiple van Hove singularities in Nb-substituted CsV$_3$Sb$_5$
along the M - K path. 
(a-c) Tomographic sections of the van Hove singularities of 
the $\alpha$-, $\beta$-, and $\gamma$-bands. 
Panels 1-6 correspond to the energy-momentum slices of Fig.~\ref{FigMCD}a 
at $k_y = -0.38$, $-0.28$, $-0.19$, $-0.09$, $0$, and $+0.09$~\AA$^{-1}$, respectively. 
Panels 1 and 5 respectively cross the K and M high-symmetry points. 
The van Hove singularities vHS1 and vHS2 of the $\alpha$- and $\beta$- bands have hole-like dispersions along K-M-K and electron-like dispersions along $\Gamma$ - M (blue and orange lines in a,b). The curvatures are opposite for the van Hove singularity of the 
$\gamma$-band (red lines in panel c). 
The data are obtained with a photon energy of 170~eV, corresponding to 
$k_z \approx 10\times G_{001}$. (Representation similar to Fig.~3 in 
Ref.~\onlinecite{Kang2022})
}
\end{figure*}

\begin{figure*}
\includegraphics[width=0.7\textwidth]{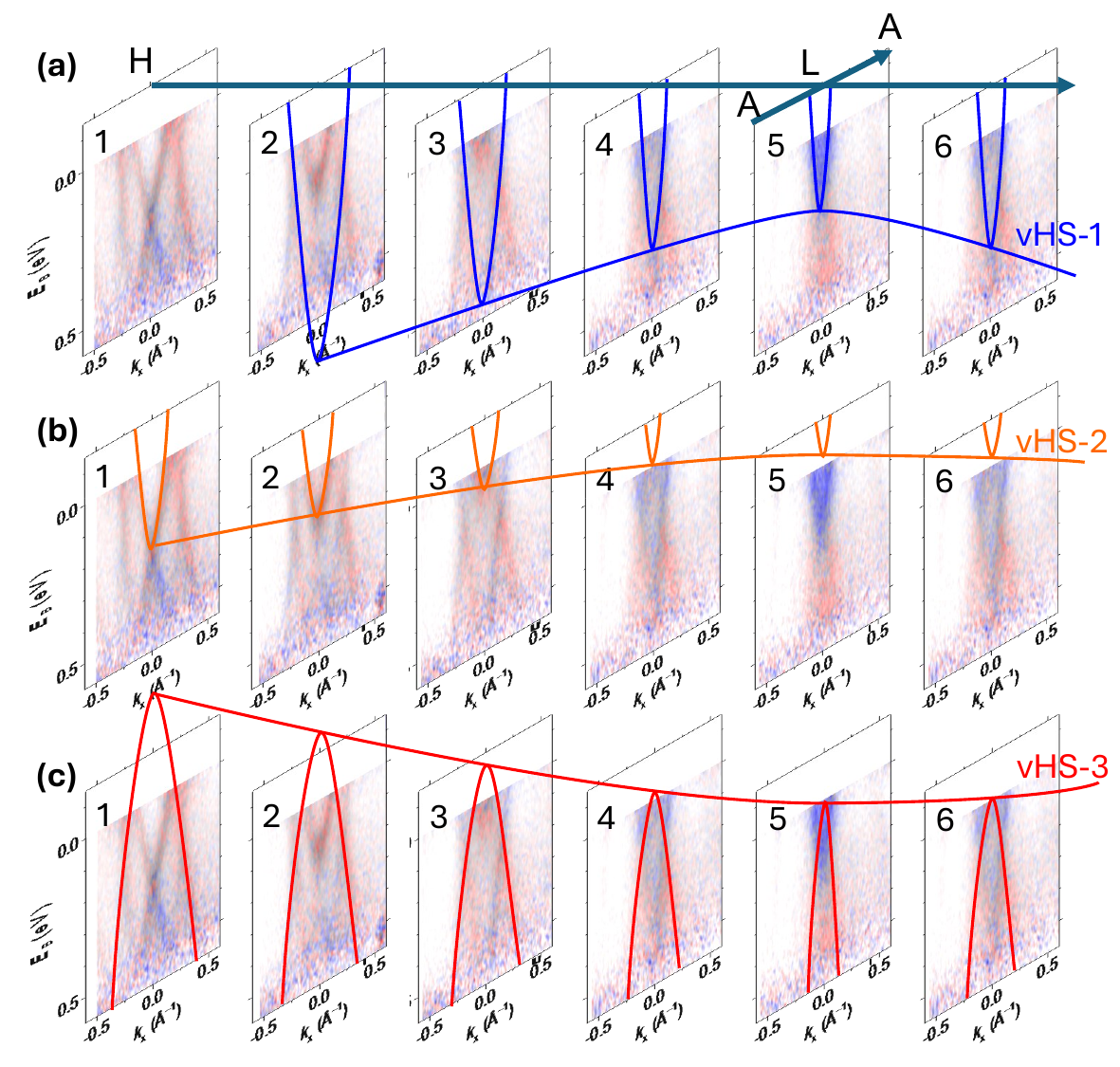}
\caption{\label{FigvHs2} 
Mapping of multiple van Hove singularities in Nb-substituted CsV$_3$Sb$_5$
along the L - H path. 
(a-c) Tomographic sections of the van Hove singularities of 
the $\alpha$-, $\beta$-, and $\gamma$-bands. 
Panels 1-6 correspond to the energy-momentum slices of Fig.~\ref{FigMCD}a 
at $k_y = -0.38$, $-0.28$, $-0.19$, $-0.09$, $0$, and $+0.09$~\AA$^{-1}$, respectively. 
Panels 1 and 5 respectively cross the H and L high-symmetry points. 
The van Hove singularities vHS1 and vHS2 of the $\alpha$- and $\beta$- bands have hole-like dispersion along H-L-H and electron-like dispersion along L - A (blue and orange lines in panels a and b). The curvatures are opposite for the van Hove singularity of the 
$\gamma$-band (red lines in panel c). 
The data are obtained with a photon energy of 155~eV, corresponding to 
$k_z \approx 9.5\times G_{001}$. (Representation similar to Fig.~3 in 
Ref.~\onlinecite{Kang2022})
}
\end{figure*}

The large magnetic circular dichroism observed in Nb-substituted CsV$_3$Sb$_5$ provides access to detailed mapping of the MCDAD signal, particularly in the vicinity of van Hove singularities. The saddle-point behavior of the van Hove singularities vHS1, vHS2, and vHS3, associated with
the $\alpha$, $\beta$, and $\gamma$ bands near the M-point, respectively, are illustrated
in Figs.~\ref{FigvHs1} and \ref{FigvHs2}. 
According to the representation in Ref.~\onlinecite{Kang2022} of the pristine CsV$_3$Sb$_5$ kagome metal, the three van Hove singularities vHS1, vHS2, and vHS3 can be identified 
by their dispersions with opposite curvatures in two orthogonal directions along M - $\Gamma$ and M - K lines (see Fig.~\ref{FigvHs1}a-c). 
In the case of the $\alpha$ band marked in Fig.~\ref{FigvHs1}a, the bottom 
of the electron-like parabola along the M-$\Gamma$ direction (shown in 
Fig.~\ref{FigvHs1}a5) moves up in energy as it approaches the M-point 
- see Figs.~\ref{FigvHs1}a2-a5. 
Along the K-M-K directions it has a hole-like dispersion with a maximum at the M point (blue line). Along the $\Gamma$-M-$\Gamma$ direction, the $\alpha$ band shows electron-like dispersion (Figs.~\ref{FigvHs1}a5), forming a saddle point, or van Hove singularity, at the binding energy $E_B= 0.15$~eV. 
Near the M-point, the $\alpha$ band exhibits the most significant (negative) MCDAD values, suggesting that vHS1 is responsible for the emergence of orbital magnetic moments.

The $\beta$ band, marked in Fig.~\ref{FigvHs1}b, 
shows similar saddle-point behavior, albeit with smaller hole-like curvature
along the K-M-K path. Here, the van Hove singularity appears at a binding energy close to $E_F$. The electron-like dispersion along $\Gamma$-M-$\Gamma$ path only becomes visible as the K-point is approached. 
This electron-like dispersion develops into the upper branch
of the lifted Dirac point with a finite mass at the K point - see Fig.~\ref{FigvHs1}b1. 

The saddle point vHS3, which is related to the $\gamma$ band and displays the opposite curvatures, is marked in Fig.~\ref{FigvHs1}c. 
Here, the band maximum of the hole-like band dispersion 
along the $\Gamma$-M-$\Gamma$ path (see Fig.~\ref{FigvHs1}c5) 
moves up in energy as it approaches the K point, as shown in Fig.~\ref{FigvHs1}c1-c4. 

Significant MCDAD values are also observed in the A-L-H plane of the Brillouin zone, as shown in Fig.~\ref{FigvHs2}, with a distribution similar to that of the $\Gamma$-M-K plane.
The van Hove singularity vHS1 at $E_B=0.15$~eV and the associated $\alpha$ band in the L point show negative MCDAD, as shown in Fig.~\ref{FigvHs2}a5.
Conversely, the $\beta$ and $\gamma$ bands near the L point exhibit a positive MCDAD.
The van Hove singularity vHS2, formed by the $\beta$ band, lies above the Fermi level at the L point. Consequently, electron-like dispersion is observed along the A-L-A path as one approaches the H point, as shown in
Figs.~\ref{FigvHs2}b2,b3. At the H point, the electron-like parabola transforms into the upper Dirac band.

\section{Discussion}

\blue{The orbital loop current is predominantly carried by the V $3d$ orbitals. Soft X-rays excite electrons from the initial states with $3d$ character to free-electron-like states with $sp$ character. According to the Clebsch-Gordan integrals, electronic states with the largest orbital moments have the largest transition matrix elements. Therefore, we expect that for $m=+1$ ($m=-1$) light the spin-degenerate $3d_{-2}$ to $np_{-1}$ 
($3d_{+2}$ to $np_{+1}$) transition to dominate the photoemission intensity.  At high-symmetry M-points with defined orbital character, the MCDAD sign is determined by the predominant occupation of one of these initial states, which originates from the degeneracy lifting caused by the effective orbital magnetic field.}

As pointed out in Ref.~\onlinecite{Wu2021}, the coexistence of two different curvatures of the van Hove singularities near the Fermi level can be derived already at the simplest level of a minimal tight-binding model consisting of two orbitals only~\cite{Wu2021}.
This was confirmed by polarization-dependent ARPES measurements reported in Ref.~\onlinecite{Hu2023}, in which the orbital character of the states forming the van Hove singularities was identified.

The hybridization of $d_{xz}/d_{yz}$ orbitals on the kagome lattice 
results in two sets of kagome bands with opposite (odd and even) mirror eigenvalues (see Fig.~\ref{Fig1}d)~\cite{Wu2021}. 
The $\gamma$ band can be understood as the lower Dirac band with the odd mirror eigenvalue, while the $\alpha$ band corresponds to the upper Dirac band with the even mirror symmetry.
The flat band of vHS2 ($\beta$ band) was assigned to V $d_{x^2-y^2}/d_{z^2}$ orbitals~\cite{Hu2022b}. 
vHS3 arises from two sublattices and corresponds to a
mixed-sublattice type of van Hove singularity~\cite{Hu2022b,Kang2022,Wu2021},
characterized by eigenstates that are evenly distributed over two
of the three sublattices at each M point.
In contrast, vHS1 and vHS2 are orbitals of the pure type, predominantly localized in the A-site sublattice.

The aforementioned symmetry properties are different from the observed signs of the MCDAD (negative for vHS1 and positive for vHS2 and vHS3), excluding a direct link between the type of a vHS and the sign of the MCDAD. Therefore, we conclude that the tight-binding model alone cannot account for the occurrence of the observed MCDAD and the inferred orbital magnetic moments. Instead, a mechanism involving electron-electron interactions has been proposed~\cite{Li2024}. Considering that multiple vHSs with opposite mirror eigenvalues are close in energy, it has been suggested that the nearest-neighbor electron repulsion favors a ground state with coexisting loop current order and charge-bond orders~\cite{Li2024}. Here, the loop current order breaks time-reversal symmetry in the CDW phase. This model predicts a 2$\times$2 loop current order within the V plaquettes of the kagome lattice. In the reciprocal space, the loop current order predicts non-vanishing orbital moments located at the M points, which is consistent with our experimental findings.

Indeed, the large MCDAD values observed here in Nb-substituted CsV$_3$Sb$_5$ provide evidence for time-reversal symmetry breaking, which is in agreement with previous findings in the pristine CsV$_3$Sb$_5$ compound~\cite{Guo2024,Elmers2025}.
The previously observed time-reversal symmetry breaking was explained by a theoretical approach~\cite{Guo2024} that also explained the chiral loop currents. These induce a substantial and switchable magnetochiral transport asymmetry in kagome metals, when the charge-density-wave modulation is taken into account. The loop current phase then resonantly amplifies the nonlinear effects caused by electronic correlations. 

\blue{We observed the largest MCDAD asymmetries at the M points at the binding energies of the three van Hove singularities indicated in Figs.~\ref{FigvHs1} and \ref{FigvHs2}. 
This suggests that the loop current order is connected to these van Hove singularities. A complete set of possible loop current patterns in a $2\times 2$ unit cell was discussed in Ref.~\cite{Schultz2025}. 
These patterns break the translation invariance with three ordering vectors that connect the van Hove points. Regarding the orbital compositions, two different categories were discussed: those considering the phase of the nearest-neighbor hopping parameters between V sites and those considering the phase of the hopping parameters between V sites and planar Sb sites. Furthermore, in Ref.~\cite{Schultz2025}, it is suggested that the loop current ordering is connected to the low-temperature superconductivity of this material, whereby the two categories lead to different symmetries of the superconducting order parameter. Our experimental results suggest that, due to the absence of magnetic circular dichroism at the Sb states centered near the $\Gamma$ point, the loop current order is connected to the hopping parameters between the V sites, resulting in $d+id$ superconducting pairing~\cite{Schultz2025}.}


\section{Summary}

Angle-resolved photoemission spectroscopy was used to study changes in the electron band structure of Cs(V$_{1-x}$Nb$_{x})_3$Sb$_5$ with $x=0.05$, when V atoms are partially substituted by Nb atoms. We have shown that the substitution of larger, but isoelectronic, Nb atoms results in \blue{an increase of the band width}, which we attribute to increased orbital overlap. 
The van Hove singularities determine the maximum density of states at the Fermi level, but increase the gap openings at pristine compound's Dirac points.
The pronounced magnetic circular dichroism of the V $3d$ and Nb $4d$ bands \blue{strongly suggests} the predicted connection between the orbital moments and the three van Hove singularities near the Fermi level at the M points. The chemical pressure exerted by the Nb doping increases the magnetic circular dichroism compared to that of pristine CsV$_3$Sb$_5$, either due to increased orbital moments, caused by the increased electron-electron interaction, or due to increased spin-orbit coupling. Crucially, larger overall larger magnetic circular dichroism in Nb-substituted CsV$_3$Sb$_5$ enabled the detailed mapping of MCDAD, providing valuable information for future theoretical investigations into various electronic instabilities in kagome metals.



\begin{acknowledgments}
We thank Daniel Schultz, Roser Valenti and Joerg Schmalian for stimulating discussions. 
This work was funded by the Deutsche Forschungsgemeinschaft (DFG, German Research Foundation), grant no. TRR288–422213477 (projects B03, B04, and B08) and
TRR 173 268565370 (projects A02),
as well as grant no Scho341/16-1, 
and by the Federal Ministry of Education and Research (BMBF) (projects 05K22UM2, and 05K22UM4). 
We thank the Diamond Light Source for providing beamtime at beamline I09.  
\end{acknowledgments}

%

\end{document}